# Extended Abstract: A BLE and UWB Beacon-Assist Framework for Multiuser Augmented Reality Synchronization Across Multiple Devices in Shared Environments

Maitree Hirunteeyakul
*City University of Hong Kong*
Matt.H@my.cityu.edu.hk

*Abstract* - *The challenge to synchronize augmented reality (AR) across sessions/devices has been solved by relying solely on vision-feature mapping, which is suboptimal in scaling workable space and flaws under visual changes in surroundings.*

*This study implemented AR synchronization solutions utilizing location beacon technology, namely Bluetooth Low Energy (BLE) and Ultra-Wideband (UWB), to discourse scalability issues and inconsistencies in the existing AR system. The framework is bifurcated into two approaches: BLE-assist and UWB-assist AR synchronization. The BLE-assist method utilizes iBeacon technology for room context recognition, integrating with Apple's ARKit ARWorldMap and Google's ARCore Cloud Anchors. The UWB-assist solution employs precise beacon ranging capabilities fusion with the device's azimuth to establish fixed spatial reference in AR across sessions/devices.*

*Comparative evaluations show that the UWB-assist approach outperforms the BLE-assist approach in reliability across environmental variations, as it always successfully resolves virtual anchors with a near-constant latency average at 25 seconds, regardless of the physical setting changes. Conversely, the BLE-assist implementation tends to be more accurate in resolving virtual anchors with a mean of 0.02 metres in position error and within 0.03 radian in orientation error. In the UWB-assist approach, computed fixed spatial references have an average disparity of 0.04 metres and 0.11 radians in pose.*

*The UWB-assist approach is ideal for scenarios requiring consistently successful localization with acceptable accuracy. In contrast, the BLE-assist approach is more suitable when demanding finer precision in virtual anchor poses with the performance tradeoffs when the surroundings are altered, such as for destinated short-lived AR sessions.*

## INTRODUCTION

Many AR applications in the market are multi-user, involving the alignment of coordinates to create a constant spatial of the virtual world in the physical world among all sessions and devices. The existing method of synchronizing AR sessions mainly derives from visual-based environmental understanding to resolve the global coordinate system into each local AR session, e.g. ARWorldMap in Apple's ARKit and Cloud Anchors in Google's ARCore [1], conducting persistent anchoring across all sessions and devices.

However, as these APIs mainly rely on visual-feature point mapping, some threats are prone as follows:
- Scalability in larger workable space [2]
- Reliability under visual environmental changes [3]

Hence, those APIs mentioned earlier are designed to be practiced in a room-sized space among known nearby peers within a short span and are not meant to be saved or reused over time.

Extra implementation is needed to enhance the AR experience to support the digital-twin world concept, where the virtual world is steadily aligned to the physical world, accessible in scale, long-lived and durable to surrounding visual shifts.

In this study, I have developed two solutions to improve synchronization of AR sessions:

*1) BLE-assist AR synchronization*
Using iBeacon technology to broadcast location context and apply to existing AR synchronization framework for localizing in a room-based manner.

*2) UWB-assist AR synchronization*
Taking advantage of centimetre-precise ranging capability fusion with a magnetic compass heading information to establish fixed reference pose for hosting and resolving nearby anchors.

By leveraging location beacons through BLE and UWB technology, this study prototypes synchronized AR systems, facilitating a robust and scalable approach to achieve uniform virtual coordinates for every user in every session. Further, with UWB-assist AR synchronization, resilience to environmental changes AR experience across dynamic spaces could be committed.

## BLE-ASSIST AR SYNCHRONIZATION

The main idea of BLE-assist AR synchronization is to break down the spacious working space, such as a building, into rooms for optimizing the AR experience with traditional spatial anchoring frameworks. This involves obtaining the current room context the user is in and coordinating spatial anchoring frameworks accordingly.

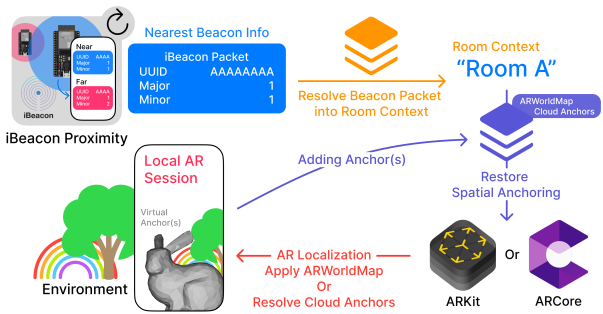

FIGURE 1. SIMPLIFIED PROCESS FOR BLE-ASSISTED AR SYNCHRONIZATION

### I. Room Context Retrieval

Our implementation will listen to iBeacon broadcasting from ESP32, which acts as beacons placed over the expansive working space to obtain room context. iBeacon standard allows devices to determine the proximity between the broadcaster (beacon) and observer (AR device), whether immediate, near or far [4].

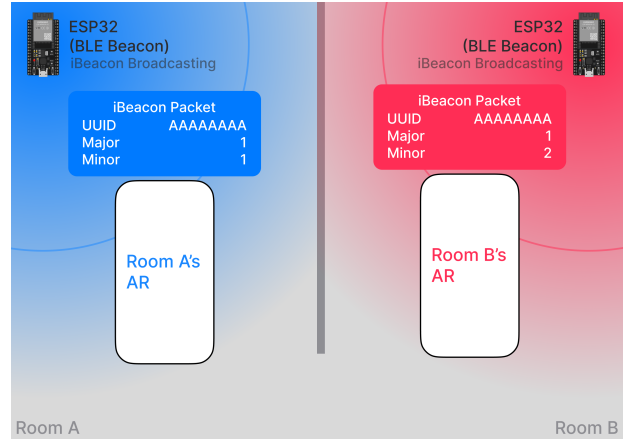

FIGURE 2. ILLUSTRATION OF iBEACON UTILIZATION FOR ROOM CONTEXT RETRIEVAL

Therefore, we can assume the nearest beacon(s) for the referencing current room, as each beacon is registered to a virtual room in a many-to-one relationship. When the current room changes, e.g., the user enters a new room, spatial anchoring updates shall be triggered. For the unflickering experience, our designed algorithm desires to minimize room context updates when the user is on the edge of multiple rooms.

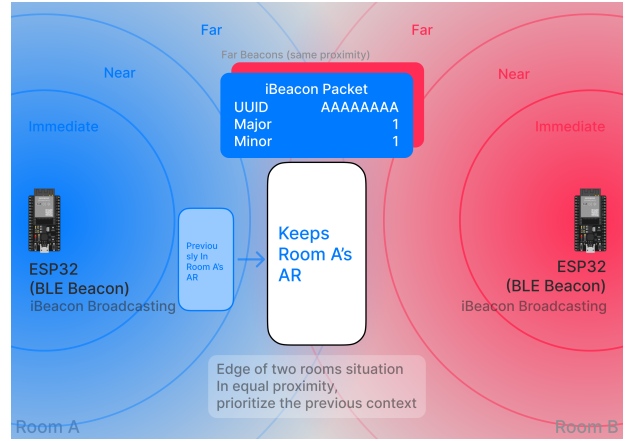

FIGURE 3. EFFORT TO MINIMIZE ROOM CONTEXT UPDATES

### II. Spatial Anchoring (Optical feature-base frameworks)

*ARWorldMap*

The common practice of ARWorldMap is to archive ongoing AR sessions from the host device as encoded data and transfer them through the network for unarchiving and localizing into receiver AR sessions [5]. Upon successful resolution, the host AR session's coordinates will be applied, and existing anchors will be passed on to receiver AR sessions.

With room context provided, our implementation can download and upload narrowed ARWorldMap according to the user location, preventing transmitting and localizing the overwhelming size of the map in an extensive workable area. However, when multiple users alter the map concurrently, saving conflict in inherited maps must be handled.

*Cloud Anchors*

In contrast to the convention of sharing maps in ARWorldMap, Cloud Anchors's practice is to host and resolve individual anchors to/from the ARCore API server [6], where each anchor is limited with a time-to-live (TTL) up to one year [7].

When a new anchor is added to any local AR session, we can host the anchor through ARCore API and record the returned anchor identifier pair with the current room context in our shared database. Once entering a new room, our implementation can pull a list of anchor identifiers associated with the room from the shared database and resolve them in the local AR session, requesting solving only nearby anchors to avoid computing unnecessary far-away anchors.

UWB-ASSIST AR SYNCHRONIZATION

The concept of UWB-assist AR synchronization constructs the reference pose of UWB beacons and deems them stationary across sessions, then uses the beacon pose to derive relative transformation for nearby anchors, either hosting or resolving with the database.

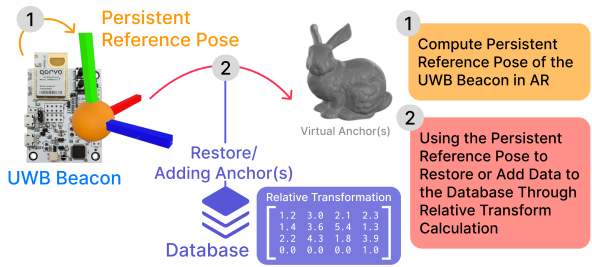

FIGURE 4. SIMPLIFIED PROCESS FOR UWB-ASSISTED AR SYNCHRONIZATION

*I. Constructing Fixed Reference Pose from UWB Beacon and Magnetic Heading*

A spatial pose consists of two parts: Position and Orientation. To establish a fixed reference pose, we must produce persistence in both position and orientation across all sessions.

*Position Reference*

This study uses DWM3001CDK, a UWB module compatible with FiRa standard [8], as a beacon. With the UWB-enabled iPhone as an AR device,

The beacon's position in AR can be obtained through the Apple Nearby Interaction framework [9], which operates Time-of-Flight UWB ranging between the AR device and the beacon.

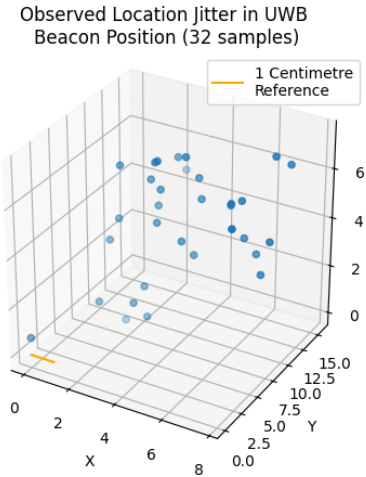

FIGURE 5. OBSERVED POSITIONS OF A STATIONARY UWB BEACON AMONG AR SESSIONS, REVEALING CENTIMETERS-PRECISION ACCURACY

*Ranging Stabilization Delay*

Per our observation, the beacon's position in AR is inconsistent during the initial ranging period.

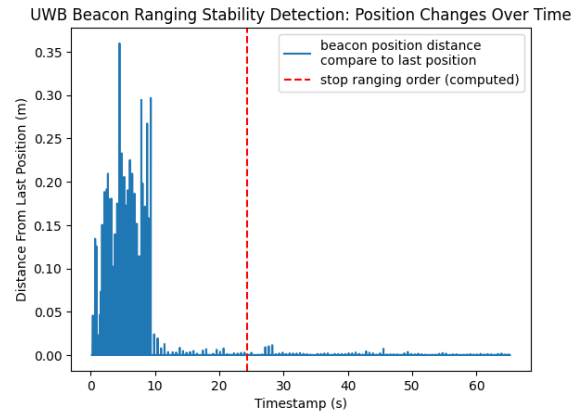

FIGURE 6. GRAPH DEPICTING THE EUCLIDEAN DISTANCE OF RANGING POSITIONS OVER TIME, WITH A CUTOFF LINE INDICATING THE ALGORITHM'S DETERMINATION OF A STABILIZED POSITION

Accordingly, we have applied a stabilization algorithm that deems a position valid once it consecutively remains within a predefined disparity threshold between each observation over a set period. Presently, ranging stabilization is the main delay in the localization of UWB-assist AR synchronization.

*Orientation Reference*

Since the UWB beacon cannot present its orientation to the device, we can refer to Earth's magnetic and gravity, as detected by the device's inertial measurement unit (IMU), to achieve persistent orientation.

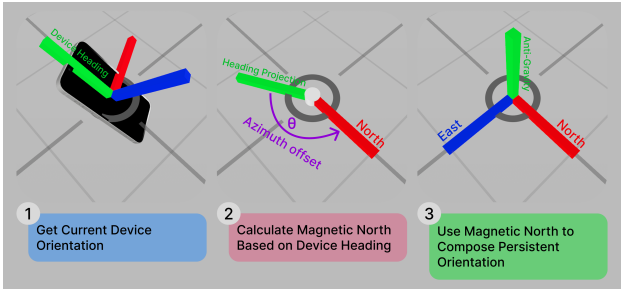

FIGURE 7. ILLUSTRATION OF THE PROCESS FOR ESTABLISHING A PERSISTENT ORIENTATION IN AR USING DEVICE ORIENTATION AND AZIMUTH

Typically, in AR frameworks, the y-axis in the AR coordinates is parallelly appointed to gravity [10], causing the xz-plane to parallel the ground. We can project the device azimuth on the ground plane from the AR camera orientation and adjust with the magnetic heading angle from IMU to create a persistent orientation where the y-axis, x-axis, and z-axis are toward the sky, magnetic north and east orderly.

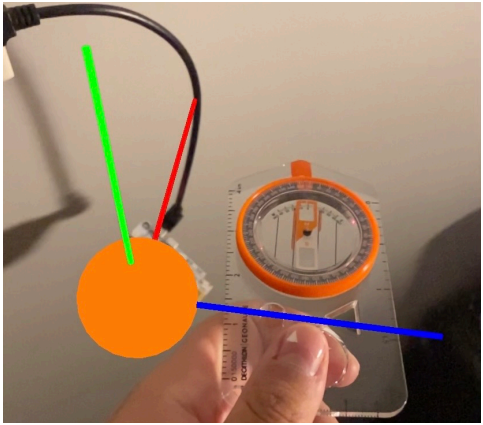

FIGURE 8. OUTCOME OF CREATING A FIXED REFERENCE POSE IN AR, DEMONSTRATED ALONGSIDE A PHYSICAL COMPASS

The orientation error is subject to local magnetic field interference [11].

## II. Relative Transformation of Anchors

After we combine persistent position and orientation to assemble a fixed reference pose, we can use this pose to host and resolve virtual anchors via relative transformation. Individual anchors' poses can be represented as homogeneous transformation matrices [12] relative to a beacon's pose and shared across AR sessions.

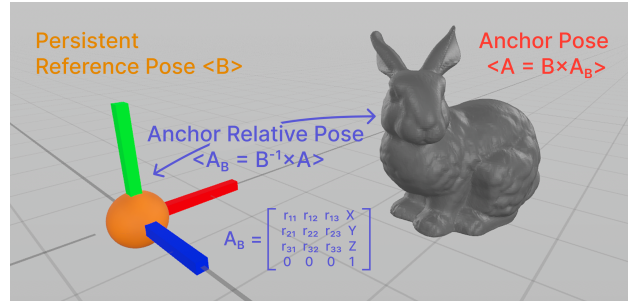

FIGURE 9. RELATIONSHIP BETWEEN FIXED REFERENCE POSE AND ANCHORS

Given that $A$ and $B$ are the transformation of an anchor and a beacon (the reference pose), respectively, and $A_B$ is the relative transformation of the anchor orienting the beacon.

*Hosting Anchors*

$$A_B = B^{-1} \times A$$

By applying the inverse of beacon pose to anchor pose using cross operation, we can get the relative transformation [13] of the anchor orienting the beacon.

*Resolving Anchors*

$$A = B \times A_B$$

The cross product of the beacon pose with the relative transformation of the anchor orienting the beacon results in the anchor pose in the world ordinates [13].

## EVALUATION

We have performed 300 trials of AR sessions with our implementation to evaluate and compare the performance of each approach.

### I. Accuracy of Localization

The error distances were calculated assuming optical AR markers' poses as the ground truth. The detected anchor pose used for analysis is relative to the markers.

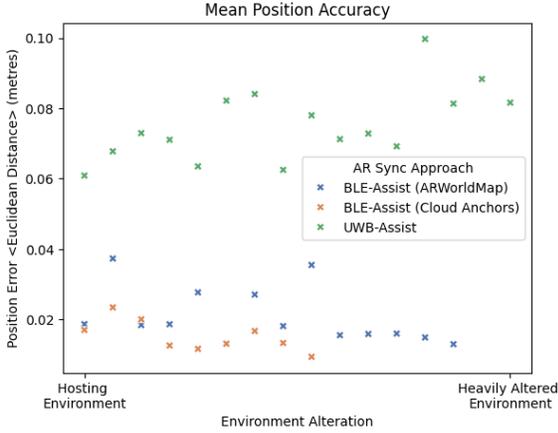

FIGURE 10. GRAPH OF MEAN POSITION ACCURACY IN ANCHORING

Since anchors are computed depending on observed reference pose in the UWB-assist approach, relative transformation may magnify the positional error by relative distance and orientation error. Therefore, to limit the interference in errors, the position of the anchor was programmatically set to the beacon.

Consequently, the positional error shown for the UWB-assist approach is the lower bound.

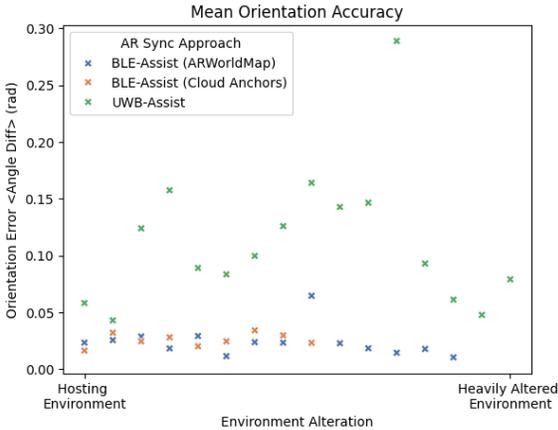

FIGURE 11. GRAPH OF MEAN ORIENTATION ACCURACY IN ANCHORING

## II. Robustness in Anchor Resolution Under Environmental Variations

In this test, we hosted the AR anchor under the hosting environment, gradually adding objects to the physical scene to simulate unexpected environmental alterations. We counted in successful resolutions in restarted AR sessions.

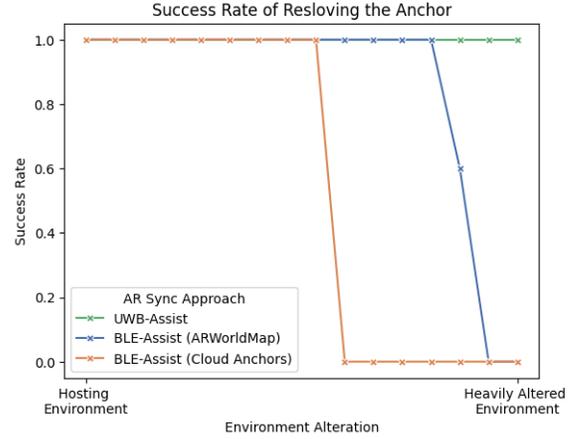

FIGURE 12. GRAPH OF SUCCESS RESOLUTION IN ANCHORING RATE

## III. Latency of Synchronization

The localization delay is measured by the time interval between the application launch and the first anchor resolution.

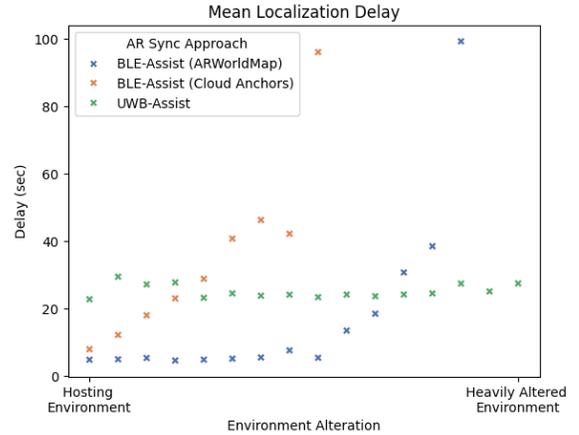

FIGURE 13. GRAPH OF MEAN LOCALIZATION DELAY

## IV. Scalability

### BLE-Assist AR Synchronization

In the ARWorldMap implementation, the scalability concerns are the map-saving conflict and the delay due to downloading and uploading the map file. On the other hand, using Cloud Anchors needs maintenance effort per the anchor's TTL restriction [7]. Moreover, per our observation, the ARCore API is not accessible under the Great Firewall for mainland Chinese users.

### UWB-Assist AR Synchronization

The scalability concern of this approach is the need for ranging. Each AR device needs to range with each beacon till finished for resolving in anchoring. However, the DWM3001CDK's UWB chip can range concurrently with up to eight devices [14]. Users may need to wait for ranging with the beacon if many devices are on the venue, growing overall localization delay.

## V. Power Consumption

On the site, two primary sources contribute to power consumption: the beacon and the AR device. The power usage of the hardware beacon increases with the number of beacons deployed and scales with the size of the workable area—meanwhile, the consumption from the AR application scales with the number of users.

TABLE 1. ELECTRICITY USAGE PER HARDWARE BEACON

| AR Sync Approach | Hardware | Power (W) | Cost Per Year (USD) |
|---|---|---|---|
| BLE-Assist | ESP32 | 0.41 | 0.61 |
| UWB-Assist | DWM3001CDK | 0.05 | 0.08 |

Price calculation is based on electricity prices for businesses in Hong Kong [15].

TABLE 2. AR APPLICATION ENERGY IMPACT

| AR Sync Approach | | Energy Impact |
|---|---|---|
| BLE-Assist | ARWorldMap | High |
| | Cloud Anchors | High |
| UWB-Assist | | High (Slightly Higher) |

The energy impact is monitored using Xcode. The result is in Apple's energy impact standard.

## VI. Deployment Cost

TABLE 3. COST PER HARDWARE BEACON

| Item | Cost Per Month (USD) | Remark |
|---|---|---|
| Backend Server | 5 | - |
| Database | 15 | - |
| Object Storage (S3) | 5 | For ARWorldMap approach only. |
| ARCore API | Free | For Cloud Anchors approach only, with API rate limits |

Above displayed cost is a rough declaration of fixed costs for prototyping with minimum requirements to keep the system up and running.

TABLE 4. COST PER HARDWARE BEACON

| AR Sync Approach | Hardware | Cost Per Beacon Setup (USD) |
|---|---|---|
| BLE-Assist | ESP32 | 17 |
| UWB-Assist | DWM3001CDK | 40 |

## DISCUSSION

As BLE-assist implementation depends on optic reference and UWB-assist relies on a radio signal and magnetic field measurement, the BLE approach is more precise in pose accuracy. In contrast, the UWB approach is worse but still in a reasonable accuracy range. Our further investigation into the accuracy of reference pose establishment in the UWB-assist implementation shows promising results.

TABLE 5. ACCURACY METRICS FOR REFERENCE POSE ESTABLISHMENT IN UWB-ASSIST AR SYNCHRONIZATION

| Metric | Mean | Max |
|---|---|---|
| Position Difference | 0.04 Metres | 0.13 Metres |
| Orientation Difference | 0.11 Radian | 0.69 Radian |

These metrics are calculated by evaluating all pairwise session combinations.

Regarding reliability, the UWB approach is ideal, consistently achieving successful localization, while BLE implementation fails to perform under heavily altered environments.

Although the localization delay of UWB-Assist AR Synchronization is not pleasant, it could be enhanced with a more effective stabilization algorithm or the ability to know when ranging results have stabilized. In addition, there is room for improving scalability at the hardware level of the beacon's UWB chip for more concurrent communication capacity.

One advantage of the UWB approach is the ability to manipulate anchoring via homogenous metric without needing an AR session, unlike in the BLE approach, where operating through existing spatial anchoring frameworks, hosting anchors is executable only through ongoing AR sessions.

## CONCLUSION

In the effort of AR synchronization in extensive areas, the BLE beacon can be used to smaller the map into rooms for local AR sessions to resolve through the existing spatial anchoring framework. In contrast, the stationary UWB beacon can establish a persistent pose as the reference for virtual anchors via relative transformation, achieving consistent anchoring among AR sessions.

Per our evaluation, the BLE-assist AR synchronization is accurate in the anchor's pose subject to inconsistency in the successfulness of localization. Meanwhile, in UWB-assist AR synchronization, the pose accuracy is lower despite the consistent success of localization. The localization delay of the BLE approach is subjective to environment variation, unlike the UWB approach, where the delay is constant. Although the average localization delay of the UWB approach is unsatisfactory, it could be improved in various parts.

In summary, the UWB-assist AR synchronization is suitable for extensive workable areas in AR when emphasizing successful anchor resolution over high precision, as its accuracy falls within a fair range. In distinction, the BLE-assist AR synchronization is ideal for short-lived AR sessions due to the inability to resolve under environment alteration, thus the benefit of higher accuracy in anchoring.

CODE AVAILABILITY

The source code for this work is publicly available at https://github.com/ARBeacon.